\documentclass[12pt]{article}

\usepackage[labelfont=bf,labelsep=space]{caption}
\usepackage{times}
\usepackage{cite}
\usepackage{graphicx}% Include figure files
\usepackage{epstopdf}
\usepackage{dcolumn}% Align table columns on decimal point
\usepackage{bm}% bold math
\usepackage{amsmath}
\usepackage{soul}

\usepackage{graphicx}

\newenvironment{sciabstract}{%
	\begin{quote} \bf}
	{\end{quote}}

\title{Quantum Coherence Tomography of Lightwave--Controlled Superconductivity} 
\author
{L. Luo$^{1\ast}$, M. Mootz$^{1,2\ast}$, J.~H.~Kang$^{3\ast}$, C. Huang$^{1}$, K. Eom$^{3}$, J.~W.~Lee$^{3}$, C. Vaswani$^{1}$, \\Y. G. Collantes$^{4}$, E. E. Hellstrom$^{4}$, I.~E.~Perakis$^{2}$, C.~B.~Eom$^{3}$ and J. Wang$^{1\dag}$
	\\
	\normalsize{$^{1}$Department of Physics and Astronomy, Iowa State University,}\\
	\normalsize{and Ames Laboratory, Ames, IA 50011 USA}\\
	\normalsize{$^{2}$Department of Physics, University of Alabama at Birmingham,}\\
	\normalsize{$^{3}$Department of Materials Science and Engineering, University of Wisconsin-Madison,}\\
	\normalsize{Madison, WI 53706, USA}\\
	\normalsize{$^{4}$Applied Superconductivity Center, National High Magnetic Field Laboratory,} \\
	\normalsize{Florida State University, Tallahassee, FL 32310, USA.}\\
	\normalsize{$^\ast$Equal contribution}\\
	\normalsize{$^\dag$To whom correspondence should be addressed; E-mail: jwang@ameslab.gov.}
}

\date{}

\begin{document} 
	
	
	\baselineskip24pt
	
	
	\maketitle

	
	\begin{sciabstract}
		Lightwave periodic driving of nearly dissipation-less currents has recently emerged as a universal control concept for both superconducting (SC) and topological electronics applications.
	While exciting progress has been made towards THz-driven superconductivity, our understanding of the interactions able to drive non-equilibrium pairing is still limited, partially due to the lack of direct measurements of high-order correlation functions. Such measurements would exceed conventional single-particle spectroscopies and perturbative responses to fully characterize quantum states far--from--equilibrium. 		
	Particularly, sensing of the exotic collective modes that would uniquely characterize lightwave-driven SC coherence, in a way analogous to the Meissner effect, is very challenging but much needed.
	Here we report the discovery of lightwave-controlled superconductivity via {\em parametric} time-periodic driving of the strongly--coupled bands in iron-based superconductors (FeSCs) by a unique phase--amplitude collective mode assisted by %inversion-symmetry-breaking 
	broken--symmetry THz supercurrents. We are able to measure non--perturbative, high--order correlations in this strongly--driven superconductivity by  
	separating the THz multi-dimensional coherent spectra (THz--MDCS) into conventional pump--probe, 
	Higgs collective mode, and pronounced bi--Higgs frequency sideband peaks with highly nonlinear field dependence. 
	We attribute the drastic transition in the coherent spectra to parametric excitation of time--dependent pseudo--spin canting states modulated by a phase--amplitude collective mode that manifests as a strongly nonlinear shift 
	%in the 
	%dominant THz--MDCS peaks, 
	from 
	%the Higgs mode frequency 
	$\omega_\mathrm{Higgs}$ to 2$\omega_\mathrm{Higgs}$. 
	%which forms 
%	above critical excitation.
	%A theory--experiment comparison identifies the hallmark of  
	%associates 
	%such parametrically driven superconducting state 
	%with the observation of 
	%as a strongly nonlinear shift in the 
	%dominant THz--MDCS peaks, from the Higgs mode frequency $\omega_\mathrm{H}$ to 2$\omega_\mathrm{H}$. 
	Remarkably, the latter higher--order sidebands dominate over the lower--order pump--probe and Higgs mode peaks above critical field, which indicates the breakdown of the susceptibility perturbative expansion in the parametrically-driven SC state.	 
		Correlation tomography by THz--MDCS provides opportunities for sensing of parametric quantum matter and non-equilibrium SC pairing that even processes finite center-of-mass momentum, %out of equilibrium 
	with implications for THz supercurrent acceleration to extend gigahertz quantum 
	%computing 
	circuits.
	\end{sciabstract}
	$\newline$
	
Alternating ``electromagnetic” bias, in contrast to DC bias, has shown promise to enable dynamical functionalities by terahertz (THz) modulation and control of topological-/super-currents and quantum order parameters during timescales faster than a cycle of lightwave oscillations~\cite{yang2019lightwave,vasw2020, matsunaga2014, Fausti2011,Knap2016,Mitrano2016,Budden2021,Shimano2021,Buzzi2021}.  	
THz--lightwave--accelerated SC and topological currents ~\cite{linder2015,maag2016,Reimann2018,kapetanakis,Luo, matsunaga2014, caval,Dienst2011,yang2019lightwave,Shimano2019} have revealed exotic quantum dynamics, e.~g., harmonic modes ~\cite{yang2019lightwave,vaswani2019discovery,linder2015} and gapless quantum fluid states \cite{yang2018} forbidden by equilibrium SC pairing symmetry, or light-induced Weyl and Dirac nodes~\cite{Luo,vasw2020}. 
However, 
%full characterization of the far--from--equilibrium quantum states accessed by strong and coherent THz--light--wave driving has proved extremely challenging. 
%due to  competing  contributions to the experimental signals.
high--order correlation characteristics far exceeding the known two-photon light coupling to superconductors are hidden in conventional spectroscopy signals and perturbative responses, where a mixture of multiple excitation pathways contribute to the same low-order responses~\cite{Cea2016,Murotani}. 
%light--induced high--order correlations are hidden in conventional optical spectroscopy data due to  multiple excitation pathways contributing to the measured nonlinear responses.  
%One compelling solution 
%xperimental characterization of non-equilibrium quantum states is possible if we can 
%is to identify unique signals arising from the collective modes of the underlying SC order parameters in the non-equilibrium quantum states. 
%Transitions between different driven phases, by tuning a control parameter, would then manifest themselves 
%through the observation of drastic changes in such collective mode signals.
%Close to equilibrium, the collective modes that uniquely characterize a superconducting state, analogous to the Meissner effect,  
A compelling solution to sensing SC coherence
of light--driven states 
is to be able to identify unambiguously their collective modes ~\cite{matsunaga2014,Krull2016,hybrid-higgs,Giorgianni2019,Benfatto2019,Chu2020,Schwarz2020}. The dominant collective excitations of the equilibrium SC phase range from amplitude fluctuations  (Higgs mode) to fluctuations of interband phase differences (Leggett mode) of the SC order parameter.
Although amplitude modes have been observed close to equilibrium when external DC ~\cite{Moor2017,Shimano2019,Puviani2020} and AC~\cite{yang2019lightwave,vaswani2019discovery} fields break inversion symmetry (IS), here we show that the phase coherent dynamics of the SC order parameters can parametrically drive quantum states, yet-to-be-observed, accessed by strong THz coherent two--pulse excitations. These states are characterized by distinct phase-amplitude collective modes  
%from those of the
%equilibrium SC state 
arising from strong light-induced couplings between the amplitude and phase channels. 
%Full characterization of the more exotic high--order correlations in a light--induced 
%superconducting state  is still elusive.  
%Importantly, the {\em control} over such light--induced correlations and excitations enable their applications to quantum information, sensing and device technologies. 

	\begin{figure}
		\begin{center}
			\includegraphics[width=160mm]{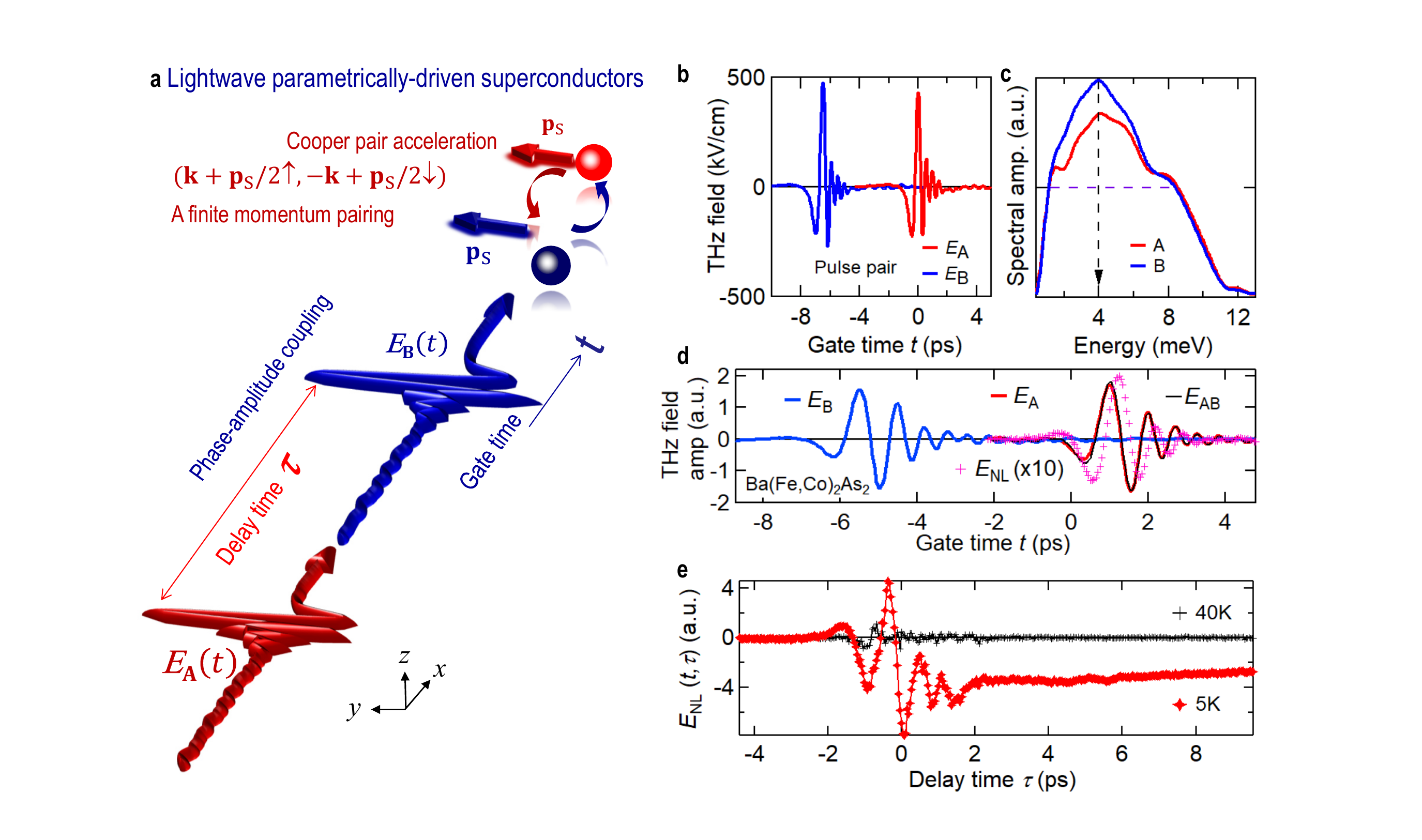}
		\end{center}
		\noindent {\textbf{Fig. 1. Terahertz (THz) multidimensional coherent spectroscopy of lightwave accelerated  non--equilibrium superfluid states in FeSCs.} 
			%up to one megavolt/cm.} 
			$\bf{a}$, Schematics of lightwave supercurrent generation, coherent control and detection of the parametrically-driven SC state via two phase-locked THz pulses in our experiment.    
			$\bf{b}$, Temporal waveforms of the nearly single-cycle THz pulse--pair used in the experiment (red and blue lines), and $\bf{c}$, spectra of the used pulses, centered at 
			$\omega_0=4$~meV.
			$\bf{d}$, Temporal dynamics of the measured coherent nonlinear transmission $E_\mathrm{NL}(t,\tau) (\mathrm{pink})=E_\mathrm{AB}(t,\tau)(\mathrm{black})-E_\mathrm{A}(t)(\mathrm{red})-E_\mathrm{B}(t,\tau)(\mathrm{blue})$ as a function of gate time $t$ at a fixed delay time between the two pulses, $\tau=6.5$~ps, under THz driving fields of 229~kV/cm at temperature of 5~K. 
			$\bf{e}$, Temporal dynamics of the $E_\mathrm{NL}(t,\tau)$ amplitude decay 
			below (red diamond, 5~K) and above (black cross, 40~K) $T_\mathrm{c}$, as a function of pulse--pair time delay $\tau$ under THz driving fields of 333~kV/cm.
			The correlated nonlinear signal $E_\mathrm{NL}(t,\tau) (\mathrm{red})$ decays over timescales much longer than the pulse duration.}
	\end{figure}
	
	THz frequency, multi-dimensional coherent nonlinear spectroscopy (THz-MDCS) )~\cite{maag2016,shaul, Kuehn2011,Junginger2012, Nelson,Johnson2019,Mootz2022}  represents a correlation tomography tool to distinguish between different 
	%coherent nonlinear responses and identify high--order 
	many-body response functions and light--induced  collective modes in  superconductors
	under strong two--pulse THz excitation. 
	Unlike for THz--MDCS studies of semiconductors~\cite{maag2016, Kuehn2011,Junginger2012,peter}, magnets~\cite{Nelson}, and molecular crystals~\cite{Johnson2019}, Fig.~1a illustrates three distinct features of our scheme in superconductors, not explored so far.    
	%Athough THz-MDCS has been widely applied in  semiconductors, magnets and molecular systems, it has not been applied to 
	%and, thereby, not impactful in the study of 
	%any superconducting correlated systems until this work.  
	%from previously studied ones. 
	{\em First}, our approach is based on measuring {\em the phase of the supercurrent coherent nonlinear emission}, in addition to the amplitude, by using phase-resolved coherent measurements with two {\em intense} phase--locked THz pulses of {\em similar} field strengths that has not been applied on superconductors. Taking advantage of both the real time and the relative phase of the two THz fields, we separate in two-dimensional (2D) frequency space spectral peaks generated by light-induced correlations and collective mode interactions 
	from  the conventional pump--probe, four--wave--mixing, and high--harmonic--generation signals~\cite{yang2019lightwave,hybrid-higgs}. This 2D separation of spectral peaks arising from high--order nonlinear processes achieves a ``super" resolution of many-body interactions 
and collective modes in highly non-perturbative states, which is not possible with the conventional spectroscopy one-dimensional measurements of prior works~\cite{hybrid-higgs, yang2019lightwave,yang2018}.  
	{\em Second},
	%the light--induced supercurrent-carrying state entangles Anderson pseudo-spins (PSs) at different momentum states. In particular, 
	as a result of lightwave condensate acceleration by the  effective local field inside 
	a thin--film SC induced by two sub-gap THz pulses and electromagnetic propagation effects,  the Cooper pairs  ($\mathbf{k}$, $\mathbf{-k}$)
	of the equilibrium BCS state 
	experience a highly nonlinear center--of--mass momentum, $\mathbf{p}_\mathrm{S}(t)$, i.~e., SC pairing with finite center-of-mass momentum (Fig. 1a). 
Precisely, the state persists well after the two strong pulses and results in ($\mathbf{k}+\mathbf{p}_\mathrm{S}(t)/2,\mathbf{-k}+\mathbf{p}_\mathrm{S}(t)/2$) Cooper pairing, due to dynamical symmetry breaking of the centrosymmetric pairing states \cite{yang2019lightwave}.
	%While the oscillating superfluid momentum $\mathbf{p}_\mathrm{s}(t)$
	%must vanish after the pulse when the  driving field temporal profile 
	%is similar to that of the applied laser field, 
	%electromagnetic propagation 
	%leads to an effective driving field which differs from the externally applied laser field 
	%due to interference between incident and reflected light--waves inside 
	%the nonlinear SC system \cite{}. 
{\em Third}, 
%as a result of interference between incident and reflected electromagnetic lightwave propagation inside 
%the nonlinear SC system~\cite{vaswani2019discovery,Mootz2020}, 
%$\mathbf{p}_\mathrm{S}(t)$
%also acquires a DC photoinduced component lasting well after the pulses~\cite{yang2019lightwave,vaswani2019discovery}. 
%In the presence of such a long--lived DC supercurrent
%Particularly, 
% well after the pulse, 
the driven quantum state with current-flow $\propto \mathbf{p}_\mathrm{S}(t)$ controllable by two--pulse interference can host distinct collective modes that provide parametric excitation of the time-dependent quantum states (Fig. 1a), whose nonlinear interactions determine the THz--MDCS spectral profile \cite{Mootz2020}.   
%be visualized directly through
% and strong Coulomb interaction between electron and hole pockets that far exceeds the intra--band pairing interaction, as in the studied  FeSC systems. 
%a persistent 
%light--induced
%breaking of the inversion symmetry (IS) of the centosymmetric paring states, which lasts well after the pulse.  
%Second, nonlinear excitations by the two intense phase-locked THz pulses 
%generate high--order correlations and new collective modes that are a direct consequence of phase transition in the non--equilibrium 
%SC states. They evolve from the zero--momentum--pairing to a current--flowing condensate state with finite--momentum
%pairing and strong Coulomb interaction between electron and hole pockets that far exceeds the intra--band pairing interaction, as in the studied  FeSC systems. 
%Below we attribute the measured changes in the spectral profile and peaks 
%of the THz--MDCS spectra  with increasing THz pulse--pair driving 
%to the changes in the collective modes characterizing these lightwave-driven phases. 

In this Article, we reveal  a  superconducting state  parametrically driven by time--periodic  light--induced dynamics of the order parameter phase in a Ba(Fe$_{1-x}$Co$_{x}$)$_2$As$_2$ superconductor. Such parametric driving becomes important when the phase dynamics is amplified by a phase--amplitude collective mode that develops with increasing THz pulse--pair driving.
The change in the character of the SC state manifests itself via the drastic changes in the THz--MDCS spectra observed with increased two--pulse driving. 
%for the high fields used here. 
In particular, we observe 
% The transition in the THz--MDCS lineshape 
a transition in THz--MDCS, from wave--mixing peaks centered at the Higgs amplitude mode frequency 
$\omega_\mathrm{H}$ to bi--Higgs frequency sidebands at  2$\omega_\mathrm{H}$. 
Remarkably, the higher--order 2$\omega_\mathrm{H}$ peaks dominate over the lower order single--Higgs peaks and 
conventional pump--probe spectra above critical THz field strength. We attribute the non--perturbative sidebands at bi--Higgs frequencies 
to the parametrically driven SC state by the collective temporal oscillations in the phases of the $s_{\pm}$--symmetry order parameter, consistent with our quantum kinetic simulations. 
 % fully reproduce the experimentally observed non-perturbative THz-MDCS features of such driven SC states. These  simulations corroborate our
%proposed interpretation of the 2$\omega_\mathrm{H}$ sideband peaks  as arising from nonlinear interaction  between a  phase-amplitude collective mode and a quasi--particle excitation. This nonlinear process parametrically 
%drives time--dependent canting of the Anderson pseudospins from their equilibrium directions in the $s_{\pm}$ SC state, 
%due to the emergence of a long--lived phase--amplitude collective mode with $\omega_\mathrm{H}$  phase
%oscillations
%above a threshold field.  

We measured optimally Co-doped BaFe$_2$As$_2$ (Ba-122) epitaxial thin film (60~nm) with $T_\mathrm{c} \sim$23~K and lower SC gap 2$\Delta_1\sim$6.8~meV (Methods section~1.1). 
%In THz-MDCS measurements, the FeAs film is excited with two nearly single cycle THz pulses with center frequency $\sim$4~meV and broadband frequency width of $\Delta\omega\sim$2~meV (Figs.~1b-1c).
%We characterized the coherent nonlinear dynamics directly in the time domain by measuring THz-MDCS in the collinear laser pulse transmission configuration shown schematically in Fig.~1a (Methods). 
We used THz-MDCS to measure the responses to two phase-locked, nearly single-cycle THz pulses A and B of similar field strength (Fig.~1b), with central frequency $\omega_0\sim$4~meV (black arrow, Figs.~1c) and broadband frequency width of $\Delta\omega\sim$6~meV (purple dashed line, Fig.~1c) (Methods section~1.2). 
Representative time scans of these THz-MDCS experiments driven by laser fields $E_\mathrm{THz, A, B}=$229\,kV/cm, are shown in Fig.~1d.
The measured nonlinear differential emission correlated signal, $E_\mathrm{NL}(t,\tau)=E_\mathrm{AB}(t,\tau)-E_\mathrm{A}(t)-E_\mathrm{B}(t,\tau)$, was recorded 
%for the  collinear pulse geometry  of Fig.~1a 
as a function of both the gate time $t$ (Fig.~1d) and the delay time $\tau$ between the two pulses A and B (Fig.~1e).  
We note three points. 
%salient features of the measured two--pulse coherent signals. 
First, as demonstrated  by $E_\mathrm{NL}(t,\tau)$ shown  in Fig.~1d (pink cross), measured at fixed delay $\tau=6.5$~ps, the electric field in the time domain 
%, measured by a gate pulse, 
allows for simultaneous amplitude-/phase-resolved detection of the coherent nonlinear responses induced by the pulse--pair and has negligible 
%We obtain  
contributions 
%to the measured correlated  signal 
from the individual pulses. This is achieved by subtracting the individual responses, $E_\mathrm{A}(t)$ and $E_\mathrm{B}(t,\tau)$ (red and blue solid lines), from the full signal obtained in response to both phase--locked pulses, $E_\mathrm{AB}(t,\tau)$ (black solid line).
Second, the $E_\mathrm{NL}(t,\tau)$ in Fig.~1e vanishes above the SC transition temperature $T_\text{c}$, as seen by comparing the 5~K (red diamond) and 40~K traces (black cross). Third, the THz-MDCS signals persist even when the two pulses do not overlap in time, e.~g., at $\tau=6.5$~ps (Figs.~1d--1e). The long--lived correlated signal $E_\mathrm{NL}(t,\tau)$ indicates that the two sub-gap laser  excitations, centered below 2$\Delta_1$ (Fig.~1c), have generated robust supercurrent--carrying macroscopic states persisting well after the pulse. 
%Fig.~1e demonstrates the robustness of the light--induced moving condensate state, 
%which  persists for many ps in unconventional FeSCs, as witnessed by the long--lived correlated signal 
%$E_\mathrm{NL}(t,\tau)$ observed below $T_c$.
	
\begin{figure}
	\begin{center}
		\includegraphics[width=150mm,angle=-0,origin=c]{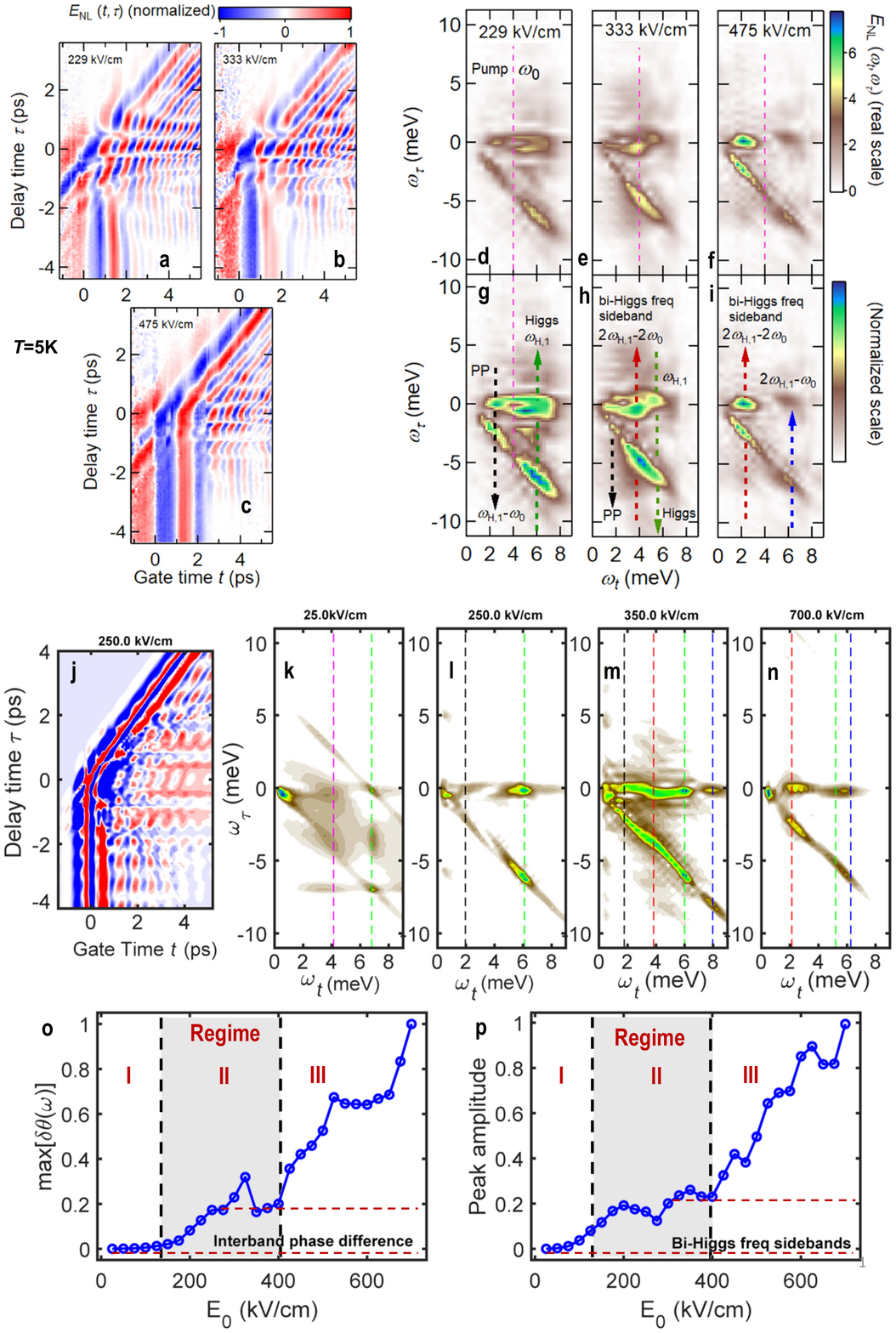}
	\end{center}
\end{figure}
\begin{figure}
	\noindent {\textbf{Fig.~2. Drastic changes of correlations and collective modes revealed in the driving electric field dependence of THz-MDCS.} 
		%up to one megavolt/cm.} 
		$\bf{a}$-$\bf{c}$, Two-dimensional (2D) false-colour plot of the measured coherent nonlinear transmission $E_\mathrm{NL}(t,\tau)$ of FeSCs superconductors at 5~K induced by  THz pump electric fields of ($\bf{a}$) 229~kV/cm, ($\bf{b}$) 333~kV/cm and ($\bf{c}$) 475~kV/cm. 
		%The THz pump-induced spectra are recorded as a function of the pump-probe delay time at 4 K. 
		$\bf{d}$-$\bf{f}$, The corresponding THz 2D coherent spectra $E_\mathrm{NL}(\omega_t,\omega_\tau)$ at 5~K for the above three pump electric fields, respectively. $\bf{g}$-$\bf{i}$, The normalized $E_\mathrm{NL}(\omega_t,\omega_\tau)$ spectra are plotted for the same pump fields to highlight the pump field-dependent evolution of the correlation peaks along the 2D frequency vector space. Peaks marked by the dashed lines are located at frequencies associated with the Higgs (green) mode and bi-Higgs frequency sideband (red and blue) consistent with the theory. $\bf{j}$, An example of calculated $E_\mathrm{NL}(t,\tau)$ as a function of gate time $t$ and delay time $\tau$ for 250~kV/cm pump field. $\bf{k}$-$\bf{n}$, 2D Fourier transform of $E_\mathrm{NL}(t,\tau)$. As predicted by the theory, dashed black (blue) lines indicate pump-probe $\omega_t=\omega_\mathrm{H,1}-\omega_0$ (bi-Higgs frequency sideband $\omega_t=2\omega_\mathrm{H,1}-\omega_0$) while IS-breaking signals at Higgs $\omega_t=\omega_\mathrm{H,1}$ (bi-Higgs frequency sideband $\omega_t=2\omega_\mathrm{H,1}-2\omega_0$) are marked by vertical dashed green (red) line; pump--probe peaks at $\omega_t=\omega_0$ are indicated by vertical dashed magenta lines. $\bf{o}$, Field-strength dependence of the dominant peak in  the spectrum of the interband phase difference $\delta\theta(\omega)$. $\bf{p}$, Field-strength dependence of the bi-Higgs frequency 
		sidebands at $2\omega_\mathrm{H,1}-\omega_0$ follows the $\delta\theta(\omega)$ behavior in (o), which identifies the importance of  light--induced time--periodic phase dynamics at the $\omega_\mathrm{H,1}$ frequency in driving a non--equilibrium SC state. Three excitation regimes are marked (main text).
		% in a way analogous to the Floquet mechanism.
	}
\end{figure}

	Figure~2 compares the 2D THz temporal profile of the coherent nonlinear signal $E_\mathrm{NL}(t,\tau)$ for relatively weak (Fig.~2a), intermediate (Fig.~2b), and strong (Fig.~2c) driving fields.
	The $E_\mathrm{NL}(t,\tau)$ dynamics reveals that pronounced coherent temporal oscillations last much longer than the temporal overlap between the two driving pulses (Fig.~1b). 
	These long--lived coherent responses generate sharp  THz-MDCS spectral peaks  visible up to $\sim$8~meV below substrate absorption (Methods). These spectra were obtained by Fourier transform of $E_\mathrm{NL}(t,\tau)$ with respect to both $t$ (frequency $\omega_t$) and $\tau$ (frequency $\omega_\tau$) (Figs.~2d--2f).
	We observe multiple distinguishing and well-defined resonances with unique lineshapes 
	that drastically change with increasing field strength. 
These $E_\mathrm{NL}(\omega_t,\omega_\tau)$ spectra differ strongly from the conventional ones measured, e.~g., in semiconductors~\cite{Kuehn2011,Junginger2012,maag2016}. 
	In the latter uncorrelated systems, 
	peaks are observable at multiples of the THz driving pulse frequency $ \omega_0\sim$4~meV (magenta dashed line), 
	as expected in the case of a rigid excitation energy bandgap.   
The observed peaks in FeSCs are much narrower than the excitation pulse width $\Delta\omega$ (Fig.~1c). This result implies that 
	$E_\mathrm{NL}(t,\tau)$ oscillates with the frequencies of SC collective mode excitations
	that lie within the $\Delta\omega$ of the few-cycle driving pulses, with the width of the THz--MDCS spectral peaks determined by the SC mode damping and not by $\Delta\omega$ of the driving pulses.
	
	The normalized $E_\mathrm{NL}(\omega_t,\omega_\tau)$ experimental spectra shown in Figs.~2g--2i 
	%allow for direct 
	 visualize %tion in two--dimensional frequency space of 
	  nonlinear couplings  of SC collective mode resonances and their field-dependences.  
	For the weaker pump field of $E_0=229$~kV/cm in Fig.~2g, the THz--MDCS spectrum shows four dominant peaks. Intriguingly, the two strongest peaks are located at the higher frequencies, roughly $(6,0)$~meV and $(6,-6)$~meV, with the weaker peaks at the lower frequencies, slightly below $(2,0)$~meV and $(2,-2)$~meV. This observation is in strong contrast to the expectation from conventional harmonic generation that high--order nonlinear signals should be weaker than lower--order ones.  
	%In particular, 
	%the higher frequency peaks at $(6,0)$=3$(2,0)$ and $(6,-6)=3 (2,-2)$
	%must come from higher--order nonlinear processes as compared to the peaks at 
	%$(2,0)$ and $(2,-2)$ and therefore one would expect them to be weaker 
	%rather than dominant as in the experiment.
	Such a reversal 
	%dominance 
	of 
	%higher--order 
	coherent nonlinear signal strengths indicates a breakdown of a susceptibility perturbative expansion around the SC equilibrium state. 
	For the intermediate field of $E_0=333$~kV/cm (Fig.~2h), the THz-MDCS spectrum shows several peaks close to each other (red and green lines), centered at new frequencies $\sim(5,0)$~meV and $(5,-5)$~meV which exhibit the similar non-perturbative behavior with dominant high order THz-MDCS spectral peaks. 
	The spectral profile changes again with increasing THz driving: four peaks are observable in the THz-MDCS spectrum for the highest studied pump field of $E_0=475$~kV/cm. For such high fields, the two strongest THz--MDCS peaks are roughly located at $(2.3,0)$~meV and $(2.3,-2.3)$~meV, while two weaker peaks become detectable at higher frequencies, at $(6.2,0)$~meV and $(6.2,-6.2)$~meV (Fig.~2i). These high field peaks should be distinguished from the low field ones at similar frequencies (Fig.~2g), as the latter have red-shifted with increasing field due to the SC gap reduction. The evolution of the well-defined MDCS spectral peaks reflect the emergence of different collective modes with increasing driving field, which characterize the transition to different non-equilibrium SC states.      
	
	We use three principles to classify the observed peaks in $(\omega_t,\omega_\tau)$ space. 
	% in two--dimensional frequency space  $(\omega_t,\omega_\tau)$.  
	 First, we introduce 
	frequency vectors characterizing the two pulses A and B,  $\omega_\mathrm{A}=(\omega_0\pm\Delta\omega,0)$ and $\omega_\mathrm{B}=(\omega_0\pm\Delta\omega,-\omega_0\mp\Delta\omega)$, which are centered around $\omega_0\sim$4~meV (black arrow, Fig.~1c). The corresponding ``time vectors"   $t^\prime=(t,\tau)$ allow us to represent the  driving electric fields of Fig.~1b in the form  ${\bf E}_\mathrm{A}(t^\prime) \sin(\omega_\mathrm{A} t^\prime)$ and ${\bf E}_\mathrm{B}(t^\prime) \sin(\omega_\mathrm{B} t^\prime)$, respectively. 
	Second, the non-equilibrium SC state driven by the above pulse--pair is characterized by a quenched asymptotic value of the time--evolved SC order parameter, 
	which defines the Higgs frequencies $\omega_{\mathrm{H},i}=2\Delta_{\infty,i}$, where $i=1$ ($i=2$) denotes the hole (electron) pocket of the FeAs bandstructure. The above Higgs mode frequencies decrease from their equilibrium values 
	of $2 \Delta_{0,i}$ 
	%\sim 2 \omega_0$ 
	with increasing field, which leads to a redshift of the THz--MDCS 
	spectral features observed in 
	Figs.~2g--2i.
	Note that we only probe the lower Higgs mode, $\omega_{\mathrm{H},1}\sim 6.8$~meV, while the higher Higgs frequency, $\omega_{\mathrm{H},2}\sim 19$~meV, lies outside of the measured spectral range (Fig.~1c). 
	%As shown theoretically, $\omega_{\mathrm{H},2}$ is also strongly damped due to the asymmetry between the electron and hole pockets in FeAs~\cite{hybrid-higgs}.  
	Third, the THz pulses drive the  Anderson pseudospin oscillators~\cite{yang2019lightwave,Mootz2022} at the different momenta ${\bf k}$ (Methods section~1.4), whose dynamics is dominated by frequencies $\sim \omega_\mathrm{H,i;A}=(\omega_\mathrm{H,i},0)$ and $\sim \omega_\mathrm{H,i;B}=(\omega_\mathrm{H,i},-\omega_\mathrm{H,i})$, i.~e., field-dependent Higgs and quasi-particle pair excitations, or  $\sim 2 \omega_\mathrm{A,B}$, i.~e., quasi-particle excitations driven at the laser frequency. 
	
	To identify which nonlinear process generates each peak measured in Figs.~2g--2i, 
	% interpret  the origin of the observed changes in the THz--MDCS spectral profile,  
	we use our quantum kinetic simulations (Methods section~1.3) and the above three principles. Light--wave propagation inside a SC thin film geometry determines the effective driving field $E(t)=E_\mathrm{THz}(t)-\frac{\mu_0 c}{2n}J(t)$, which is obtained 
from Maxwell's equations~\cite{Mootz2020}
and differs from the applied field 
$E_\mathrm{THz}(t)$ ($n$ is the refractive index).
This effective field drives the nonlinear supercurrent $J(t)$, described  self-consistently by solving the  gauge-invariant SC Bloch equations~\cite{Mootz2020,hybrid-higgs,Mootz2022}  (Methods section~1.3) for a 3-pocket SC model with strong electron--hole pocket interaction $U$ far exceeding the intra--band pairing interaction. 
 Using the above results, we 
 simulate directly the $E_\mathrm{NL}(t,\tau)$ temporal dynamics measured in the experiment (Fig.~2j as an example) and then obtain the
	$E_\mathrm{NL}(\omega_t,\omega_\tau)$ spectra (Figs.~2k--2n). These simulations are fully consistent with the observed drastic change in the THz--MDCS spectra, where non-perturbative spectral peaks emerging with increasing field, as shown in Figs.~2g--2i, are indicative of a transition to  light-driven SC states with different, emergent collective modes. 
	
	We elaborate the above quantum state transition by using three different excitation regimes, marked in Figs.~2o (red arrows) as field-strength dependence of the interband phase difference $\delta\theta(\omega)$ peak: I, perturbative regime; II, the non-perturbative state with dominant Higgs amplitude mode; regime III, the parametrically-driven SC state determined by phase--amplitude collective mode. We first examine the perturbative susceptibility regime I, where the Higgs frequency $\omega_\text{H,1}$ remains close to its equilibrium value, $2\Delta_1\sim6.8$~meV, similar to the ``rigid" excitation energy gap in semiconductors.
 The simulated THz-MDCS spectrum (Fig.~2k) then shows several peaks (Table~1, Methods) splitting along the $\omega_\tau$ vertical axis, at $\omega_t=\omega_0$ (dashed magenta line) and $\omega_t=\omega_\text{H,1}$ (dashed green line). 
	The conventional pump--probe signals are observed at $(\omega_0,-\omega_0)$ and $(\omega_0,0)$ in Fig.~2k,  generated by the familiar third-order  processes $\omega_\mathrm{A} - \omega_\mathrm{A} + \omega_\mathrm{B}$ and $\omega_\mathrm{B} - \omega_\mathrm{B} + \omega_\mathrm{A}$, respectively. Four-wave mixing signals are also observed at $(\omega_0,\omega_0)$ and $(\omega_0,-2\omega_0)$, generated by the third--order processes $2\omega_\mathrm{A} - \omega_\mathrm{B}$ and  $2\omega_\mathrm{B} - \omega_\mathrm{A}$. 
However, the perturbative behavior in this regime are inconsistent with the dominance of higher--order peaks (Fig.~2g) for the much stronger fields used in the experiment to achieve the necessary signal--to--noise ratio. 

By increasing the field strength (Figs.~2l--2n), the calculated signals 
along the $\omega_\tau$ vertical axis and at $(\omega_0,-\omega_0)$, $(\omega_0,0)$  
diminish.  
%if both pulses have  similar amplitudes.  In this strong excitation regime, 
Only peaks along $(\omega_t,0)$ and $(\omega_t,-\omega_t)$ 
are then predicted by our calculation, consistent with the experiment in Figs.~2g--2i. 
For the lower field strength of 250~kV/cm (Fig.~2l), our calculated THz-MDCS spectrum shows two weak peaks at $\omega_t\sim 2$~meV (black dashed line) and two strong broken--IS peaks at $\omega_t=\omega_\mathrm{H,1}\sim 6$~meV (green dashed line), similar to the experimental THz-MDCS peaks shown in Fig.~2g. 
The weak peaks at $\omega_t\sim 2$~meV
(black dashed line) arise from 
high-order difference-frequency Raman processes (PP, Table~2 in Methods), which generate 
pump--probe signals at $\omega_t=\omega_\mathrm{H,1}-\omega_0$, as observed in Figs.~2d and 2g.
The strong peaks at the Higgs frequency $\omega_t=\omega_\mathrm{H,1}\sim$6~meV 
(green dashed line)
dominate for intermediate fields up to $\sim$400~kV/cm (regime II, Fig.~2o), but  
 vanish if we neglect the electromagnetic propagation effects as discussed later. 
 The BCS ground state evolves into 
a finite--momentum--pairing SC state, which is determined by the condensate momentum 
$\mathbf{p}_\mathrm{S}$ 
%driven by the effective field $E(t)$. In this regime, 
%$\mathrm{p_S}$ 
generated by third--order nonlinear processes (Supplementary Fig.~4c, Note~4) and persisting well after the pulse. 
Higgs frequency peaks then arise from ninth--order IS breaking nonlinear processes generated by the coupling between the Higgs mode and the lightwave accelerated supercurrent $J(t)$ (IS Higgs, Table~2 in Methods).  
The superior resolution achieved for 
sensing the collective modes using THz--MDCS 
with 2D coherent excitation is far more than a static IS symmetry breaking scheme using a DC current (Supplementary Fig.~9, Note~7).

\begin{figure}
	\begin{center}
		\includegraphics[width=160mm]{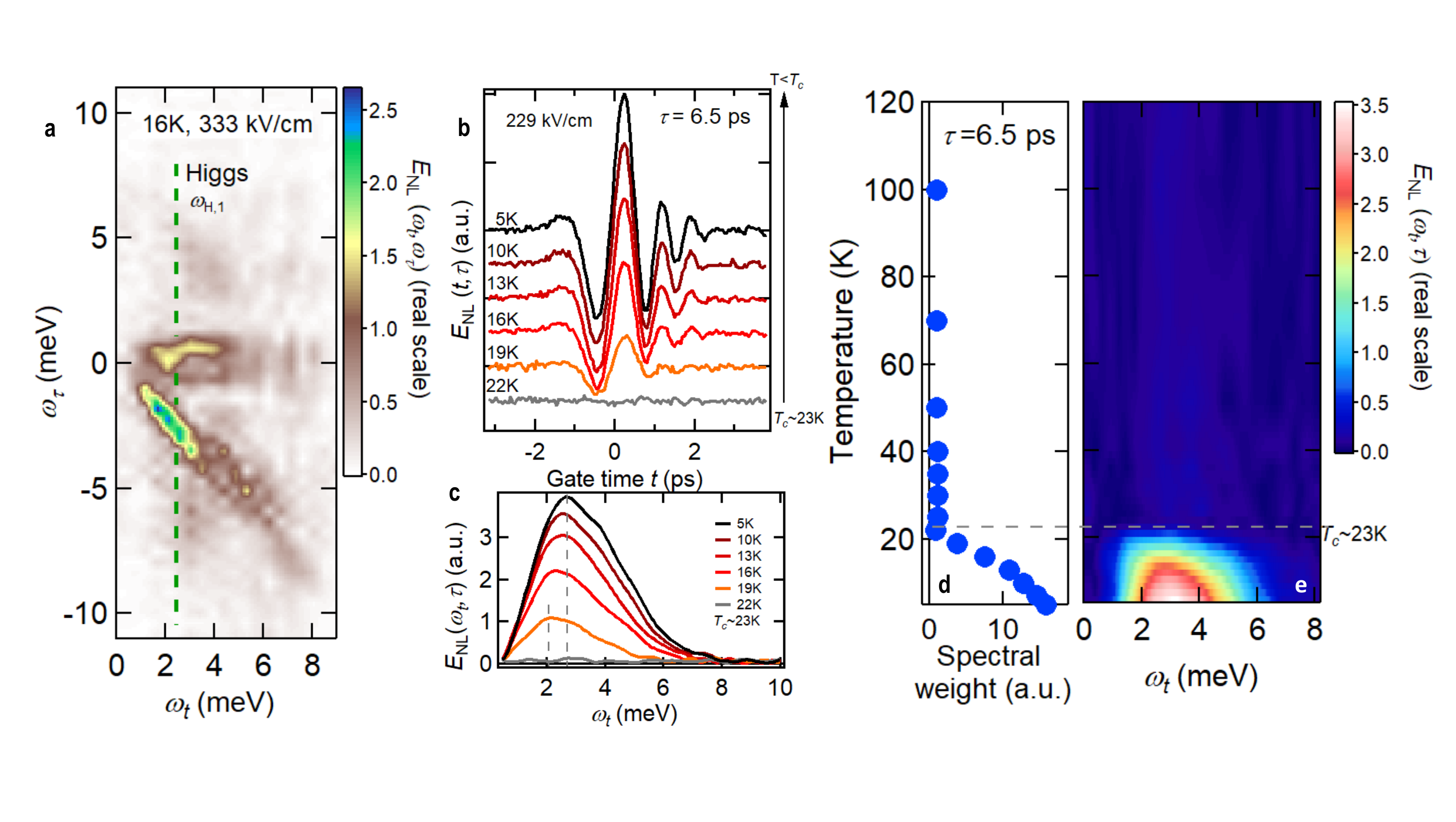}
	\end{center}
	\noindent {\textbf{Fig.~3. Temperature dependence of THz-MDCS signals.} 
		%up to one megavolt/cm.} 
		$\bf{a}$, THz-MDCS spectra $E_\mathrm{NL}(\omega_t,\omega_\tau)$ at 16~K for pump electric field 333~kV/cm.
		$\bf{b}$, Temporal profiles of two-pulse THz coherent signals $E_\mathrm{NL}(t,\tau)$ at various temperatures from 5~K to 22~K for a peak THz pump electric field of $E_{\mathrm{pump}}=229$~kV/cm and $\tau=6.5$~ps. Traces are offset for clarity. $\bf{c}$, The corresponding Fourier spectra of the coherent dynamics in ($\bf{b}$). 
		$\bf{d}$-$\bf{e}$, A 2D false-color plot of THz coherent signals ($\bf{e}$) as a function of temperature and frequency $\omega_t$ with ($\bf{d}$) integrated spectral weight at various temperatures. Dashed gray line indicates the SC transition temperature. }
\end{figure} 

For an even higher field strength of 350~kV/cm (Fig.~2m) and of 700~kV/cm (Fig.~2n), the THz--MDCS spectra change above the  excitation threshold where the order parameter phase dynamics becomes significant, as shown in Fig.~2o (regime  III). 
As discussed in Methods (section~1.4),
% and Supplementary Note~3, 
time--periodic phase deviations from their equilibrium values,  
with frequency $\omega_\mathrm{H,1}$, parametrically drive 
pseudo--spin time--dependent canting (Supplementary Fig.~4d), which coincides with a highly
nonlinear field--dependence of $\mathbf{p}_\mathrm{S}$ (Supplementary Fig.~4c). 
In this regime III, new dominant THz--MDCS peaks emerge at $\omega_t=2\omega_\mathrm{H,1}-\omega_0$ (blue dashed line), referred as to bi-Higgs frequency sideband, 
while satellite peaks are also observed at $\omega_t=2\omega_\mathrm{H,1}-2\omega_0$ (red dashed line). 
The spectral position of these emergent sideband peaks in 2D frequency space 
indicates that they arise from a second harmonic of the Higgs frequency,
$2\omega_\mathrm{H,1}$. 
Figure~2p demonstrates a threshold nonlinear behavior of these bi-Higgs frequency sideband peak strengths, which coincides 
with the development of strong phase dynamics as seen in Fig.~2o.

These theoretical prediction are fully consistent with our experimental observations in Figs.~2h and 2i. 
For the intermediate field in Fig.~2m, 
the THz--MDCS peaks at $\omega_t=2 \omega_\mathrm{H,1}-2\omega_0\sim$4 meV (red dashed line)
and $\omega_t=\omega_\mathrm{H,1}\sim$6 meV (green dashed line) 
%$\omega_t=2 \omega_\mathrm{H,1}-2\omega_0$ 
are close to each other, so they merge into a single broad resonance around $(5,0)$~meV and $(5,-5)$~meV in the calculated 2D spectrum. 
This result agrees with the measured broad, overlapping THz--MDCS peaks $\sim$5 meV in Fig.~2h, while the calculated $\omega_t=2\omega_\mathrm{H,1}-\omega_0$ peak (blue line) is not visible experimentally due to the substrate absorption. 
For the highest studied field strength (Fig.~2n), the calculated THz--MDCS signals are dominated by the bi--Higgs frequency nonlinear sidebands at $\omega_t=2\omega_\mathrm{H,1}-\omega_0\sim 6.0$~meV and $\omega_t=2\omega_\mathrm{H,1}-2\omega_0\sim 2.0$~meV. Both sidebands peaks now fall into the substrate transparency region and are clearly resolved in Fig.~2i. The emergence of these new MDCS peaks in Regime III will be discussed later as manifestations of a phase--amplitude collective mode that parametrically drive pseudo-spin canting with respect to the $s_{\pm}$--symmetry equilibrium directions. 

Figure~3 demonstrates the strong temperature dependence and redshift 
	of the observed peaks as we appoach $T_\mathrm{c}$. 
	The THz-MDCS spectrum  $E_\mathrm{NL}(\omega_t,\omega_\tau)$ at temperature 16~K is shown in Fig.~3a for the intermediate field of $E_0=333$~kV/cm, and is  compared in Fig.~2h with the spectrum at $T$=5 K for same excitation.    
	%10~K, 12~K, 14~K, 16~K, 20~K, 22~K 24~K and 30~K.  
	The broken--IS signals observed at the Higgs mode frequency $\omega_t=\omega_{\mathrm{H},1}$ red-shift with increasing temperature, from $(5,0)$~meV and $(5,-5)$~meV peaks at 5~K to broad peaks slightly below $(2.5,-2.5)$~meV and $(2.5,0)$~meV at 16~K (green line). This redshift arises from the thermal quench of the SC order parameter $2\Delta_1$ with increasing temperature. However, unlike for the case of
	THz coherent control of the order parameter demonstrated in Fig.~2h, a thermal
	quench does not produce any obvious bi--Higgs frequency THz--MDCS peaks expected $\sim$1~meV, which is indicative of the coherent origin of the latter.   
	Figures~3b and 3c show the temperature dependence of the measured differential coherent emission $E_\mathrm{NL}(t,\tau)$ and the corresponding $E_\mathrm{NL}(\omega_t,\tau)$ at a fixed pulse separation $\tau=6.5$~ps. 
	%, in the measurable frequency range $\omega_t \sim$0.5--8~meV. 
	It is clearly seen, by comparing the 5~K (black line) and 22~K (gray) traces, that, when approaching $T_\text{c}$ from below, the coherent nonlinear emissions quickly diminish and red-shift. 
	Finally, Figs.~3d--3e show a detailed plot of $E_\mathrm{NL}(\omega_t,\tau)$ up to 100~K.
	%at $\tau$=6.5\,~ps 
	%for temperatures ranging from 5~K to 120~K (Fig.~2e). 
	The integrated spectral weight shown in Fig.~3d correlates with 
%	The strong temperature dependence of the THz--MDCS coherent signals 
the SC transition at $T_\text{c}$ (gray dashed line). 
	
 \begin{figure}
 	\begin{center}
 		\includegraphics[width=170mm, origin=c]{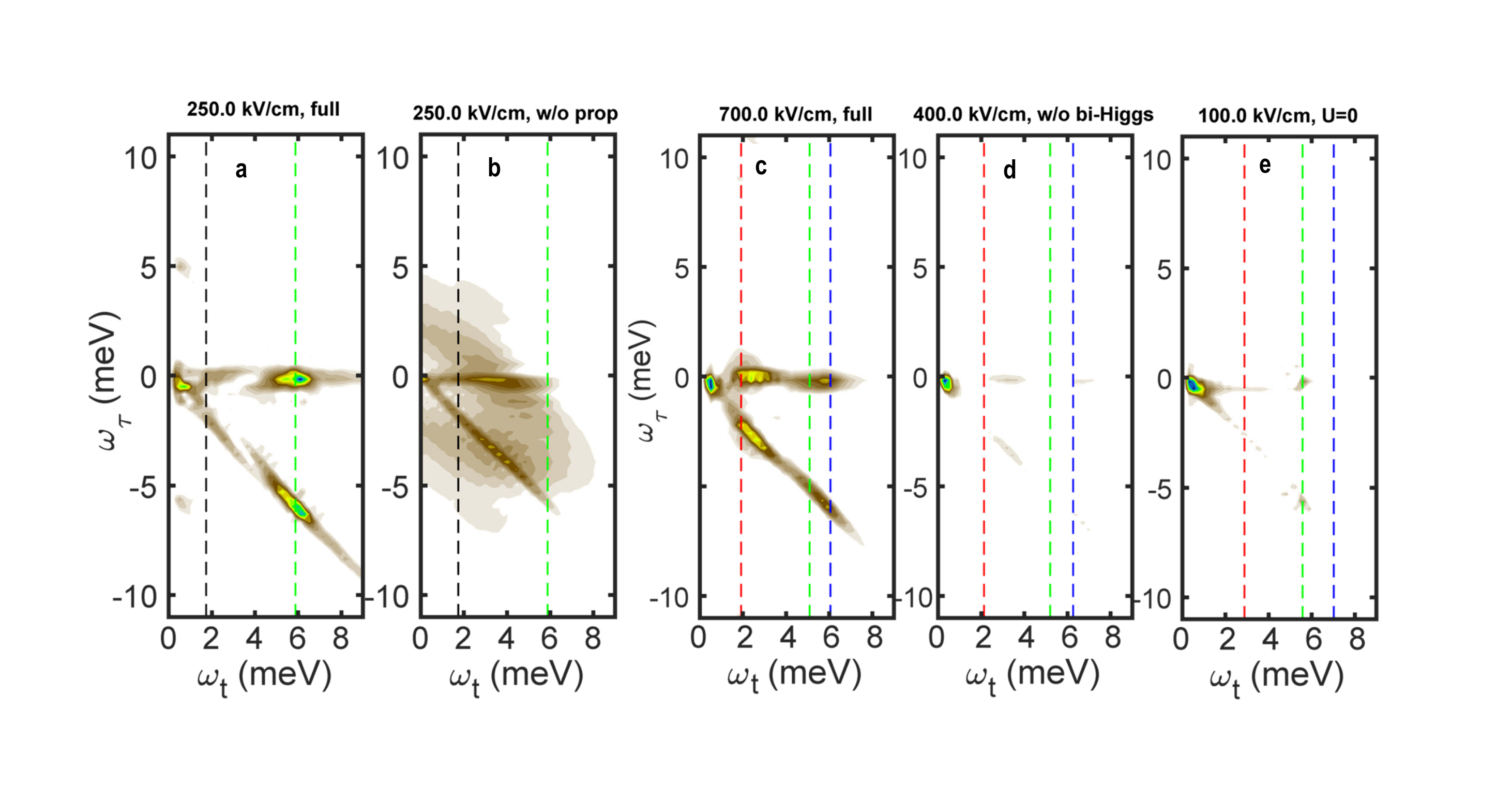}
 	\end{center}
 	\noindent {\textbf{Fig.~4. Origin of correlation and collective mode peaks in THz-MDCS signals underpinned by switch-off analysis.} 
 		$\bf{a}$-$\bf{b}$,, $E_\mathrm{NL}(\omega_t,\omega_\tau)$ for ($\bf{a}$) the full calculation with lightwave propagation ($E_0=250$~kV/cm) and for ($\bf{b}$) a calculation without propagation ($E_0=250$~kV/cm). Dashed black (green) lines indicate $\omega_t=\omega_\mathrm{H,1}-\omega_0$ ($\omega_t=\omega_\mathrm{H,1}$). The IS peaks at $\omega_t=\omega_\mathrm{H,1}$ vanish without persisting IS breaking. $\bf{c}$-$\bf{e}$, $E_\mathrm{NL}(\omega_t,\omega_\tau)$ for ($\bf{c}$) the full calculation with lightwave propagation ($E_0=700$~kV/cm), for ($\bf{d}$) a calculation without phase--amplitude coupling
 		($E_0=400$~kV/cm), and for ($\bf{e}$) a calculation without interband interaction, $U=0$ ($E_0=100$~kV/cm).  To directly compare the different THz-MDCS spectra, the field strengths of the different calculations are chosen such that $\omega_\mathrm{H,1}$ are comparable, $\omega_\mathrm{H,1}\sim 5.0$~meV. Dashed blue lines indicate $\omega_t=2\omega_\mathrm{H,1}-\omega_0$, while IS-breaking signals at Higgs $\omega_t=\omega_\mathrm{H,1}$ (bi-Higgs $\omega_t=2\omega_\mathrm{H,1}-2\omega_0$) are marked by vertical dashed green (red) lines. Note that the bi-Higgs frequency sideband  peaks are strongly suppressed without phase--amplitude coupling or without interband interaction. }
 \end{figure}
 
 Figure~4 offers more insight into the 
 physical mechanism behind  the observed transition in the THz--MDCS spectra with increasing field. First, 
 we compare the spectra $E_\mathrm{NL}(\omega_t,\omega_\tau)$ for a field strength of $250$~kV/cm between (i) the full calculation that includes electromagnetic propagation and interference effects leading to slowly decaying $\mathbf{p}_\mathrm{S}(t)$ after the pulse (Fig.~4a), and (ii) a calculation where these effects are switched off, in which case  $\mathbf{p}_\mathrm{S}(t)$ oscillates during the THz pulse and vanishes afterwards (Fig.~4b). 
 In the latter case, the $\omega_t=\omega_\mathrm{H,1}$ peak vanishes in Fig.~4b (green dashed line), 
 and the THz-MDCS spectrum is dominated by broad pump--probe (PP) peaks at $\omega_t =\omega_0\sim$4~meV,
 %that is 
 similar to 
 %the low-field result in 
 Fig.~2k.  
 This result suggests that the peaks at $\omega_\text{H,1}$ dominating the PP peaks in nonlinear regime II, as in Fig.~2g (data) and Fig.~4a (theory),
 provide coherent sensing of non-perturbative Higgs collective modes underpinning the finite--momentum--pairing SC phase (Supplementary Note~7) different from the ground state in perturbative regime I. 
 Next, we turn to the  transition from Higgs to dominant bi-Higgs signals at $\omega_t=2\omega_\mathrm{H,1}-\omega_0$. We  
 associate this transition with the development of a time--dependent 
 canted pseudo--spin state, which is  
 parametrically--driven by amplified phase dynamics at frequency $\omega_\mathrm{H,1}$. 
 The nonlinear enhancement  of this phase dynamics 
 in regime III originates from the development of a phase--amplitude collective mode at $\omega_\mathrm{H,1}$ (Figs.~2o and 2p).
 This  phase--amplitude mode interacts with a quasi--particle 
 excitation, both with energies $\sim \omega_\mathrm{H,1}$, which 
 amplifies the THz--MDCS sideband peaks at frequencies  
 $\sim 2 \omega_\mathrm{H,1}$
 (Fig.~2n), at the expense of the 
 Higgs mode peak at  $\omega_\mathrm{H,1}$
 which dominates in regime II.

We attribute the drastic change in the THz--MDCS spectra with the non-perturbative emergence of the bi-Higgs frequency sideband in Regime III over the Higgs peak to parametric excitation of time--dependent pseudo--spin canting modulation by a phase--amplitude collective mode (Methods sections 1.4 and 1.5). 
The latter collective mode develops above critical excitation and leads to 
pronounced phase dynamics at the Higgs frequency $\omega_\text{H,1}$ (Supplementary Figs.~4a and 4b).
To further corroborate the transition from Higgs amplitude to phase--amplitude collective mode, we compare in Figs.~4c and 4d the spectra $E_\mathrm{NL}(\omega_t,\omega_\tau)$ obtained from the full calculation for 700.0~kV/cm driving
	with those  obtained by turning off the 
pseudo--spin canting 
driven by the $\omega_\mathrm{H,1}$ time--periodic  phase fluctuations 
around the 
$s_{\pm}$ equilibrium state
(Supplementary Figs.~4d and 6, Note~4). 
Our 
formulation of
the
gauge--invariant 
SC Bloch equations 
in terms of two coupled pseudo--spin 
nonlinear oscillators, presented in 
the Methods section~1.4, shows that non--adiabatic pseudo--spin 
canting is parametrically driven with time--dependent strength 
	$\sim |\Delta_1|^2 \sin(2 \delta \theta)$.
	The threshold nonlinear field dependence of this coupling, 
shown in Supplementary Fig.~4d, 
leads to the strong 
field dependence of the $\sim 2\omega_\mathrm{H,1}$
sideband (Fig.~2p). 
	By comparing Figs.~4c and.~4d, we see that the signals at frequencies $\omega_t=2\omega_\mathrm{H,1}-\omega_0$ (blue dashed line) and $\omega_t=2\omega_\mathrm{H,1}-2\omega_0$ (red dashed line) are absent 
when the order parameter phase 
can be approximated by its	equilibrium 
value.  
	We also compare the full result with a calculation without interband Coulomb interaction between the electron and hole pockets (Fig.~4e), which diminishes the bi-Higgs frequency signals.
If the inter--band Coulomb coupling  
	exceeds the intra--band pairing interaction, the  Leggett mode phase oscillations lie well within the quasi--particle continuum   (regime I), so they are overdamped 
(Fig.~2o). 
Above critical THz driving (regime III), however, the THz--modulated superfluid density of strongly--Coulomb--coupled 
electron and hole pockets (Methods section 1.5 and Supplementary Note~4)
	enhances the nonlinear coupling between the phase--amplitude collective mode  ($\omega_\mathrm{H,1}$) and 
	quasi--particle excitations ($\omega_\mathrm{H,1}$) in the parametrically-driven SC state that processes supercurrent $J(t)$--oscillations at $2 \omega_\mathrm{H,1}$ visualized as the unique bi-Higgs sidebands in the THz-MDCS spectra (Fig. 2i).

	\section*{Acknowledgments}
	THz spectroscopy work was supported by National Science Foundation 1905981 (M.M.).
	THz MDCS instrument was supported by the W.M. Keck Foundation (initial design and commission) and by the U.S. Department of Energy, Office of Science, National Quantum Information Science Research Centers, Superconducting Quantum Materials and Systems Center (SQMS) under the contract No. DE-AC02-07CH11359 (upgrade for improved cryogenic operation).
	THz spectral analysis was supported by the Ames Laboratory, the US Department of Energy, Office of Science,
	Basic Energy Sciences, Materials Science and Engineering Division under contract No. DEAC0207CH11358 (L.L.). 
	The work at UW-Madison (synthesis and characterizations of epitaxial thin films) was supported by the US Department of Energy (DOE), Office of Science, Office of Basic Energy Sciences (BES), under award number DE-FG02-06ER46327.
	Modeling work at the University of Alabama, Birmingham was supported by the US Department of Energy under contract \# DE-SC0019137 and was made possible in part by a grant for high performance computing resources and technical support from the Alabama Supercomputer Authority.

	%\end{document}\end{document}
	% For your review copy (i.e., the file you initially send in for
	% evaluation), you can use the {figure} environment and the
	% \includegraphics command to stream your figures into the text, placing
	% all figures at the end.  For the final, revised manuscript for
	% acceptance and production, however, PostScript or other graphics
	% should not be streamed into your compliled file.  Instead, set
	% captions as simple paragraphs (with a \noindent tag), setting them
	% off from the rest of the text with a \clearpage as shown  below, and
	% submit figures as separate files according to the Art Department's
	% instructions.

	\clearpage
	
	%\noindent {\bf Fig. 1.} Please do not use figure environments to set
	%up your figures in the final (post-peer-review) draft, do not include graphics in your
	%source code, and do not cite figures in the text using \LaTeX\
	%\verb+\ref+ commands.  Instead, simply refer to the figure numbers in
	%the text per {\it Science\/} style, and include the list of captions at
	%the end of the document, coded as ordinary paragraphs as shown in the
	%\texttt{scifile.tex} template file.  Your actual figure files should
	%be submitted separately.
	
\end{document}